\documentclass[sigconf]{acmart}

\usepackage{lineno}
\usepackage{algorithm}
\usepackage{algpseudocode}
\usepackage{float}
\usepackage{placeins}
\usepackage{balance} 

\makeatletter
\providecommand{\@LN@col}[1]{}
\providecommand{\@LN}[2]{}
\makeatother

\copyrightyear{2027}
\acmYear{2027}
\setcopyright{cc}
\setcctype{by-nc-nd}
\acmConference[WSDM '27]{Proceedings of the Twentieth ACM International Conference on Web Search and Data Mining}{February 15--19, 2027}{Hong Kong, Hong Kong}
\acmBooktitle{Proceedings of the Twentieth ACM International Conference on Web Search and Data Mining (WSDM '27), February 15--19, 2027, Hong Kong, Hong Kong}
\acmDOI{10.1145/3862322.3863721}
\acmISBN{979-8-4007-3117-4/2027/02}

\usepackage{graphicx}
\graphicspath{{figures/}}

\usepackage{amsmath}
\usepackage{booktabs}
\usepackage{multirow}
\usepackage{xcolor}
\usepackage{enumitem}

\title{\textsc{CodeHID}: Learning an Addressable Hierarchical Code Index for Generative Code Retrieval}

\author{Zhen Li}
\orcid{0009-0006-1390-9717}
\affiliation{%
  \institution{School of Informatics\\Xiamen University}
  \city{Xiamen}
  \state{Fujian}
  \country{China}
}
\email{zhenli0104@stu.xmu.edu.cn}

\author{Yuhong Chen}
\orcid{0009-0000-9073-7038}
\affiliation{%
  \institution{Key Laboratory of Multimedia Trusted Perception and Efficient Computing, Ministry of Education of China, Xiamen University}
  \city{Xiamen}
  \state{Fujian}
  \country{China}
}
\email{yhchen2320@163.com}

\author{Wenhao Xu}
\orcid{0009-0006-5204-4322}
\affiliation{%
  \institution{School of Informatics\\Xiamen University}
  \city{Xiamen}
  \state{Fujian}
  \country{China}
}
\email{wenhaoxu2026@163.com}

\author{Xiaodong Li}
\orcid{0000-0002-5985-9886}
\affiliation{%
  \institution{School of Informatics\\Xiamen University}
  \city{Xiamen}
  \state{Fujian}
  \country{China}
}
\email{xdli@xmu.edu.cn}

\author{Hui Li}
\correspondingauthor
\orcid{0000-0001-9139-3855}
\affiliation{%
  \institution{School of Informatics\\Xiamen University}
  \city{Xiamen}
  \state{Fujian}
  \country{China}
}
\email{hui@xmu.edu.cn}

\renewcommand{\shortauthors}{Zhen Li, Yuhong Chen, Wenhao Xu, Xiaodong Li, \& Hui Li}

\ccsdesc[500]{Information systems~Specialized information retrieval}
\ccsdesc[500]{Software and its engineering~Automatic programming}

\keywords{code retrieval; generative code retrieval; code modeling}

\begin{document}

\begin{abstract}
Code retrieval models have predominantly relied on a flat matching paradigm that treats code snippets as independent candidates, making them less capable of distinguishing similar code candidates. Generative retrieval offers a solution by constructing a learnable index over the code corpus, guiding the retriever to better understand how code candidates are semantically organized and addressed. However, naively applying generative retrieval in the code retrieval task may result in operating over an identifier space whose prefixes do not correspond to meaningful code-semantic regions. In this paper, we propose \textsc{CodeHID}, a generative code retrieval framework that reformulates the code retrieval task from flat candidate matching to coarse-to-fine semantic address generation. \textsc{CodeHID} relies on two core components. First, Pseudo-Neighbor Guided DocID Learning constructs a globally static hierarchical index by applying multi-level residual quantization and $k$-nearest-neighbor pseudo-supervision, ensuring that semantically related code snippets share prefixes while preserving target-level separability. Second, Dual-Phase DocID Generation Guidance reliably navigates this fixed index by combining training-side ranking enhancements, using hard negatives and rank distillation, with inference-side candidate constraints and prefix-aware decoding. Extensive experiments on CoSQA and ProCQA benchmarks demonstrate that \textsc{CodeHID} outperforms existing sparse retrieval, pre-trained code models, dense code retrieval, and generative retrieval baselines by a large margin in most cases, achieving particularly strong improvements in rank-one retrieval metrics. We provide the implementation at: \url{https://github.com/KDEGroup/CodeHID}.

\end{abstract}

\maketitle

\section{Introduction}

\begin{figure}[t]
    \centering
    \includegraphics[width=0.97\columnwidth]{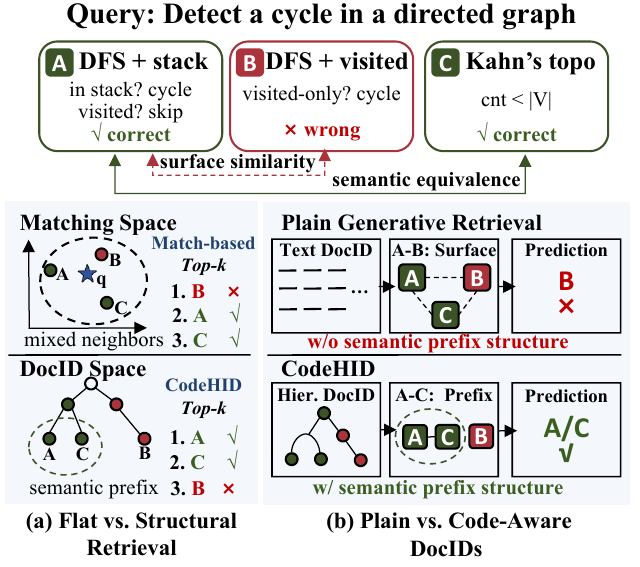}
    \vspace{-10pt}
    \caption{
    Limitations of flat matching and plain generative retrieval for code retrieval.
    (a) Flat matching may favor surface-similar but structurally incorrect code B, while structurally valid candidates A and C better match the query intent. (b) Plain textual DocIDs can inherit this surface bias, whereas \textsc{CodeHID} organizes candidates with code-aware hierarchical prefixes to retrieve semantically correct code more reliably.
    }
    \label{fig:head}
\end{figure}

When a developer issues a natural-language query for code, what should a retriever actually do?
Most methods learn a matching function between the query and candidate code snippets~\cite{GraziaP23,XieLDZW24}, encoding both into representations and ranking candidates by query-code similarity. 
This flat matching paradigm has achieved strong empirical performance~\cite{FengGTDFGS0LJZ20,GuoLDW0022,LiGSQZYQJCD22,ZhangATDNR0X24,LiuMJSXZY25,GaoZLYZCZ025,BuiYJ21,DuSWSHZ21,DongHZZWZJL25}. However, it treats each candidate as an independent text-like unit, providing limited explicit modeling of code-specific structures such as identifier relations, API dependencies, control flow, and data flow. As illustrated in Fig.~\ref{fig:head}(a), such one-step similarity ranking lacks code-aware organization of the target space, making retrieval susceptible to surface-level cues.

Recent structure-aware code models attempt to address this limitation by incorporating data flow, identifier semantics, AST information, code paths, or dependency relations into representation learning and retrieval~\cite{GuoRLFT0ZDSFTDC21,GuoLDW0022,WangWWWZLWL22,FengLWWXWL25}.
Nevertheless, most of them still use structural information as auxiliary evidence for pairwise matching.
\textbf{The retrieval candidates remain a flat set of independent snippets, and the code index itself is not organized as a learnable, fine-grained, and addressable semantic structure.}
As a result, existing methods may capture useful structural cues, but they are unable to understand relations between candidates, making them less capable of distinguishing similar code.

Motivated by the benefits of relation-aware representation learning~\cite{ChenWCDW23,WuLLCBW23}, we argue that \textbf{code retrieval should not only learn how a query matches individual snippets, but also how the code corpus itself is semantically organized and addressed}.
Code snippets often share recurring functionalities, API conventions, data-processing routines, control-flow templates, and implementation idioms~\cite{WeiL17,BuiYJ21}.
These regularities naturally induce a hierarchical organization: snippets can be grouped by functional intent, separated by programming patterns or API choices, and finally distinguished by fine-grained implementation details.
From this perspective, code retrieval should therefore move beyond one-step ranking over a flat candidate set toward semantic addressing over a structured index, where a query progressively locates its target. 

Generative retrieval~\cite{Tay00NBM000GSCM22} offers a natural mechanism for this goal. 
Unlike conventional retrieval with an external index, generative retrieval predicts document identifiers directly, making the identifier space a learnable interface between queries and retrieval targets~\cite{SiSCCZ0000G24,WangHWMWCXCZL0022,XuT0YLZL25,LiuZWJXZ00026}. 
However, as shown in Fig.~\ref{fig:head}(b), plain generative retrieval may still follow surface-driven paths when identifier prefixes do not correspond to meaningful code-semantic regions.
Effective generative code retrieval therefore requires hierarchical semantic identifiers, enabling coarse-to-fine address generation in which shallow prefixes locate broad semantic regions and deeper tokens distinguish implementations. 

The above observation raises two key challenges: 
(1) First, the DocID space should be semantic, hierarchical, discriminative, and globally static: it should organize the corpus into prefix-addressable semantic regions so that retrieval can proceed from coarse code semantics to a specific target, while keeping each target stably addressable within the index.
Prevalent reconstruction-oriented discrete identifiers~\cite{LiJZZZZD25} may compress code representations without preserving retrieval-oriented neighborhoods or meaningful prefix-sharing structures.
(2) Reliable query-to-DocID generation remains difficult even with hierarchical identifiers.
Because DocIDs are generated autoregressively, early prefix errors may lead decoding into incorrect branches and hinder recovery~\cite{Zeng0Z24,Zeng0JSWZ24,ZhangL0DLC24}.
Moreover, standard sequence-generation objectives reproduce the target DocID without explicitly discriminating it from other valid candidate identifiers~\cite{Tay00NBM000GSCM22,WangHWMWCXCZL0022,0002DW23}.
Thus, generative code retrieval requires both a hierarchical semantic address space and query-conditioned guidance for selecting the correct generation path.

To address these problems, we propose \textsc{CodeHID}, a generative code retrieval framework that learns and queries an addressable hierarchical code index.
Rather than treating the code corpus as a flat collection of candidates, \textsc{CodeHID} organizes code snippets into a globally static hierarchical DocID space and retrieves targets by generating their semantic addresses. 
\textsc{CodeHID} consists of two key components.
First, Pseudo-Neighbor Guided DocID Learning constructs the addressable hierarchical code index by quantizing code representations into multi-level discrete identifiers and using semantic-neighborhood pseudo labels to regularize prefix sharing.
This component encourages semantically related code snippets to share prefixes at appropriate depths while preserving the separability of implementation-distinct targets.
Second, Dual-Phase DocID Generation Guidance improves query-to-DocID generation for navigating this fixed identifier space.
It strengthens discrimination among similar DocID candidates during training and stabilizes prefix-level path selection during inference. 
Together, these components reframe code retrieval from pairwise query-code matching over a flat candidate set to coarse-to-fine semantic address generation over a learned hierarchical code index.

Our main contributions are summarized as follows:
\begin{itemize}[leftmargin=*,topsep=2pt]
    \item We propose \textsc{CodeHID} that reformulates code retrieval as generative semantic addressing over a learned hierarchical code index, moving beyond flat query-code matching.

    \item We propose Pseudo-Neighbor Guided DocID Learning, which converts continuous code-neighborhood relations into discrete prefix-sharing supervision to learn hierarchical DocIDs with semantic sharing and target-level separability.

    \item We introduce Dual-Phase DocID Generation Guidance, combining training-side ranking enhancement and inference-side prefix guidance for reliable query-to-DocID generation.

    \item Experiments on two code retrieval benchmarks, together with ablation and rank-one discrimination analyses, demonstrate the effectiveness of \textsc{CodeHID} and its components.
\end{itemize}

\section{Preliminaries}

Code retrieval aims to identify relevant code snippets from a corpus given a natural language query.
Let $\mathcal{C}=\{c_1,\dots,c_N\}$ denote a code corpus, where $c_i$ is the $i$-th code snippet, and let $q\in\mathcal{Q}$ denote a natural language query.
A widely adopted formulation~\cite{GraziaP23,XieLDZW24} embeds queries and code snippets into a shared continuous representation space and retrieves candidates according to their matching scores:
\begin{equation}
\label{eq:sim_retrieval}
    \hat{c}
    =
    \arg\max_{c\in\mathcal{C}}
    \operatorname{sim}(q,c),
\end{equation}
where $\operatorname{sim}(q,c)$ measures the semantic similarity between query $q$ and code snippet $c$.

Although effective, this matching-based formulation reduces retrieval to a one-step comparison in a continuous space.
Such a formulation provides limited explicit control over how code snippets are organized, distinguished, and retrieved, especially when code semantics involve structured signals such as identifiers, APIs, control dependencies, and data flow~\citep{GuoRLFT0ZDSFTDC21}.
Instead of treating retrieval as similarity-based ranking over independent candidates, we reformulate code retrieval as query-conditioned generation over a structured identifier space, i.e., \emph{generative code retrieval}.

Specifically, we assign each code snippet $c_i$ a globally static hierarchical document identifier, denoted as:
\begin{equation}
\label{eq:docid_definition}
    d_i
    =
    \left(d_i^{(1)}, d_i^{(2)}, \dots, d_i^{(L)}\right),
\end{equation}
where $d_i^{(\ell)}$ denotes the $\ell$-th token of the DocID and $L$ is the DocID length.
The DocID is determined offline by the code snippet itself and remains fixed during retrieval.
This design separates the construction of the target identifier space from query-time generation, allowing code snippets to be retrieved through their discrete hierarchical identifiers.

Let $\mathcal{D}$ denote the set of valid DocIDs induced by the code corpus.
Given a query $q$, generative code retrieval aims to generate the DocID corresponding to the target code snippet:
\begin{equation}
\label{eq:docid_objective}
    \hat{d}
    =
    \arg\max_{d\in\mathcal{D}}
    \log p_\theta(d\mid q),
\end{equation}
where $p_\theta(d\mid q)$ denotes the probability of generating DocID $d$ conditioned on query $q$.
The generated DocID $\hat{d}$ is then mapped back to its corresponding code snippet through an offline DocID-to-code index.

Since a DocID is represented as a discrete token sequence, its generation probability can be factorized autoregressively:
\begin{equation}
\label{eq:docid_ar}
    p_\theta(d\mid q)
    =
    \prod_{\ell=1}^{L}
    p_\theta\left(d^{(\ell)}\mid q,d^{(<\ell)}\right),
\end{equation}
where $d^{(<\ell)}$ denotes the previously generated DocID prefix.
Under this formulation, code retrieval is transformed from one-step similarity ranking in a continuous embedding space into coarse-to-fine generation over hierarchical identifiers.
\section{Our Method:  \textsc{CodeHID}}

\begin{figure*}[t]
    \centering
    \includegraphics[width=0.97\textwidth]{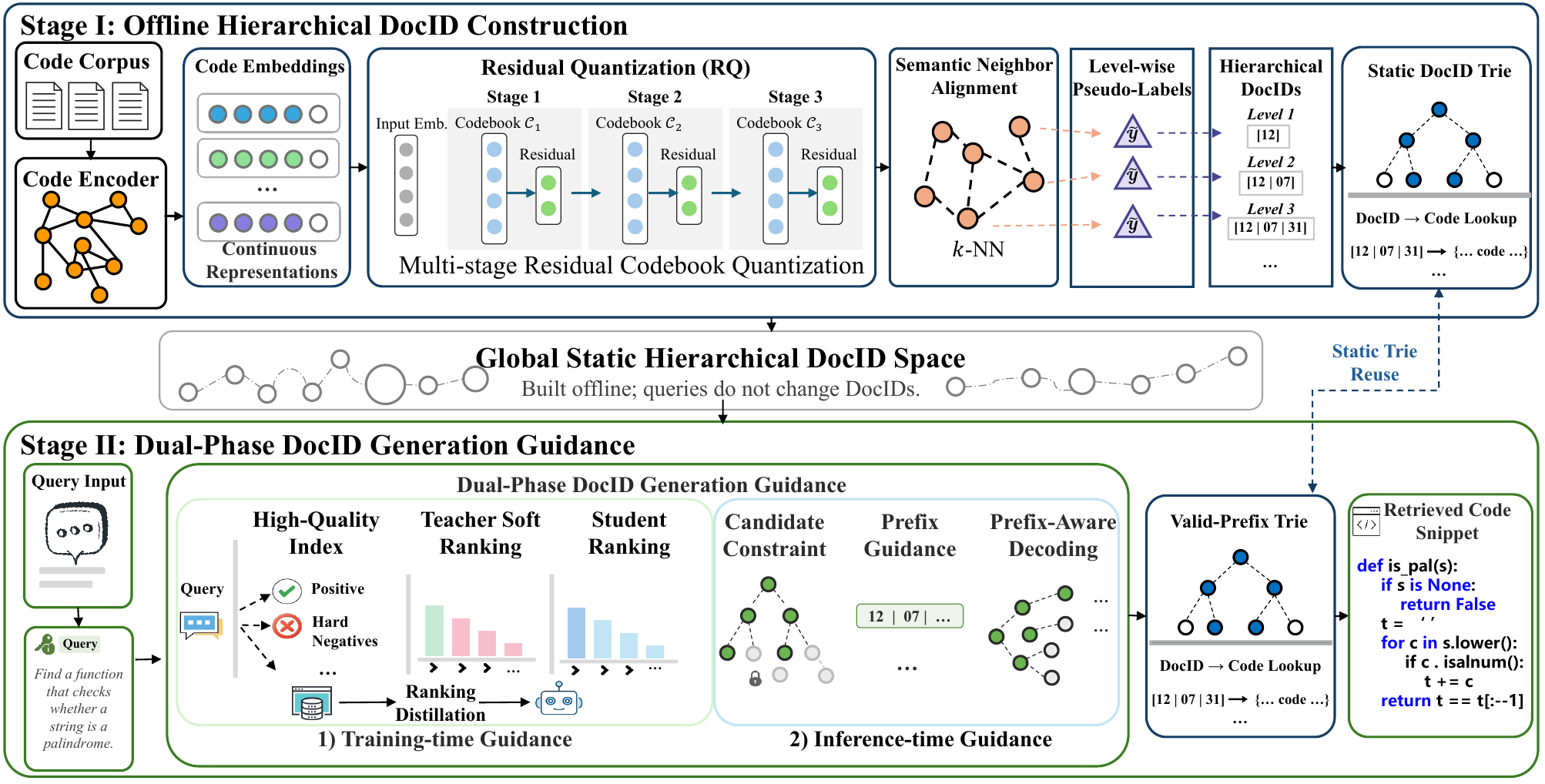}
    \vspace{-10pt}
    \caption{
    Overview of \textsc{CodeHID}. Stage I learns a globally static, code-aware hierarchical DocID space through Pseudo-Neighbor Guided DocID Learning, where residual quantization produces multi-level identifiers and semantic-neighbor pseudo labels regularize prefix sharing. Stage II queries this fixed identifier space through Dual-Phase DocID Generation Guidance, combining training-side ranking signals with inference-side candidate constraints, prefix guidance, and prefix-aware decoding. 
    }
    \label{fig:framework}
\end{figure*}

Fig.~\ref{fig:framework} depicts the overall framework of \textsc{CodeHID}. 
Rather than relying solely on one-step similarity ranking over a flat candidate set, \textsc{CodeHID} learns and queries an addressable, code-aware hierarchical index, instantiated as a globally static DocID space with an offline identifier for each code snippet.

This framework addresses two coupled challenges:
(1) how to learn a code-aware hierarchical
index that organizes code snippets through meaningful
prefix structures, and (2) how to reliably navigate this
fixed index within a query-conditioned candidate space.
Accordingly, \textsc{CodeHID} comprises two components:
Pseudo-Neighbor Guided DocID Learning, which translates
continuous code-neighborhood relations into discrete
prefix-sharing supervision, and Dual-Phase DocID
Generation Guidance, which enhances candidate
discrimination during training and guides inference by
retrieving top-$B$ candidates using $R(q,d)$, constructing a query-specific trie, and performing constrained autoregressive generation with Prefix Scoring and Query Scoring. 
The following subsections detail these components.

\subsection{Pseudo-Neighbor Guided DocID Learning}
\label{sec:rqvae_knn}

Pseudo-Neighbor Guided DocID Learning introduces \texorpdfstring{$k$}{k}-nearest-neighbor (\texorpdfstring{$k$}{k}-NN) pseudo-neighbor supervision on top of RQ-VAE~\cite{ZeghidourLOST22}, a prevalent quantization architecture adopted by \textsc{CodeHID} to iteratively quantize continuous inputs as stacked discrete codes in a coarse-to-fine manner.
To be specific, code representations are quantized into multi-level discrete codes for hierarchical DocID construction, while a \texorpdfstring{$k$}{k}-NN graph~\cite{DongCL11} in the continuous semantic space is used to identify pseudo-neighbor pairs and further derive level-wise pseudo labels.
These labels guide semantically neighboring code snippets to share DocID prefixes at appropriate depths, while discouraging excessive prefix sharing among non-neighboring samples.

To obtain pseudo-neighbor relations, a \texorpdfstring{$k$}{k}-NN graph is first built over code snippets in the continuous semantic space produced by a pre-trained code encoder.
Let $s_{ij}$ be the semantic similarity between code snippets $c_i$ and $c_j$.
The adjacency relation is defined as:
\begin{equation}
\label{eq:knn_adj}
A_{ij}
=
\begin{cases}
1, & j \in \operatorname{TopK}_{j\neq i}(s_{ij}),\\
0, & \text{otherwise}.
\end{cases}
\end{equation}
Here, $A_{ij}=1$ indicates that $c_j$ belongs to the nearest-neighbor set of $c_i$.
This adjacency relation provides the pseudo-neighbor basis for constructing level-wise pseudo labels.

\texorpdfstring{$k$}{k}-NN adjacency identifies semantic neighbors, but hierarchical DocID learning also requires a level-specific notion of prefix sharing.
Accordingly, neighborhood relations are coupled with level-wise similarity thresholds to derive prefix-sharing pseudo labels:
\begin{equation}
\label{eq:knn_pseudo_label}
\tilde{y}_{ij}^{(\ell)}
=
A_{ij}\cdot \mathbf{1}[s_{ij}\ge \tau_\ell].
\end{equation}
Here, $\tau_1<\tau_2<\cdots<\tau_L$ are level-wise similarity thresholds. 
A positive pseudo label $\tilde{y}_{ij}^{(\ell)}=1$ means that samples $i$ and $j$ are encouraged to share the first $\ell$ DocID levels.
Increasing thresholds imposes stricter semantic consistency on deeper prefixes.

The pseudo labels define the target structure of hierarchical prefix sharing.
To connect this target with the prefix-sharing pattern induced by the model, prefix-sharing probabilities are computed from the soft codeword assignments in RQ-VAE:
\begin{equation}
\label{eq:soft_assignment}
p_i^{(\ell)}(m)
=
\operatorname{softmax}_m
\left(
-\frac{
\left\|r_i^{(\ell-1)}-e_m^{(\ell)}\right\|_2^2
}{\gamma}
\right),
\end{equation}
where $r_i^{(\ell-1)}$ is the residual representation before quantization at level $\ell$, $e_m^{(\ell)}$ is the $m$-th codeword in the level-$\ell$ codebook, and $\gamma$ is the temperature parameter.

Based on the soft assignments, the probability that samples $i$ and $j$ share the first $\ell$ levels of their DocID prefixes is defined as:
\begin{equation}
\label{eq:prefix_share_prob}
\pi_{ij}^{(\ell)}
=
\prod_{u=1}^{\ell}
\left(
\sum_{m=1}^{M}
p_i^{(u)}(m)p_j^{(u)}(m)
\right),
\end{equation}
where $M$ denotes the codebook size.
The inner summation measures the probability that the two samples select the same codeword at level $u$, and the product accumulates such consistency from level $1$ to level $\ell$.
Thus, $\pi_{ij}^{(\ell)}$ can be viewed as the predicted prefix-sharing probability.

With $\pi_{ij}^{(\ell)}$, prefix-sharing supervision is imposed by aligning it with the level-wise pseudo labels derived from the \texorpdfstring{$k$}{k}-NN neighborhood.
The positive neighbor constraint encourages semantic neighbors to share prefixes at appropriate depths, while the negative separation constraint suppresses overly deep prefix overlap among non-neighboring samples.
They are defined as:
\begin{equation}
\label{eq:knn_loss_terms}
\left\{
\begin{aligned}
\mathcal{L}_{\mathrm{k\text{-NN}}}^{+}
&=
\frac{1}{|\mathcal{E}^{+}|}
\sum_{(i,j)\in\mathcal{E}^{+}}
\sum_{\ell=1}^{L}
w_\ell
\operatorname{BCE}
\left(
\pi_{ij}^{(\ell)},\tilde{y}_{ij}^{(\ell)}
\right),
\\[3pt]
\mathcal{L}_{\mathrm{k\text{-NN}}}^{-}
&=
\frac{1}{|\mathcal{E}^{-}|}
\sum_{(i,n)\in\mathcal{E}^{-}}
\sum_{\ell=1}^{L}
w_\ell
\operatorname{BCE}
\left(
\pi_{in}^{(\ell)},0
\right).
\end{aligned}
\right.
\end{equation}
Here, $\mathcal{E}^{+}$ contains in-batch neighboring
pairs in the $k$-NN graph, while $\mathcal{E}^{-}$ contains randomly sampled in-batch non-neighboring pairs; $w_\ell$ and $\operatorname{BCE}(\cdot,\cdot)$ denote the level-wise weight and binary cross-entropy loss, respectively. 
The positive term aligns semantic neighbors' prefix-sharing probabilities with level-wise pseudo labels, while the negative term discourages prefix sharing among
non-neighbors to preserve discriminability.

The final \texorpdfstring{$k$}{k}-NN prefix consistency objective is:
\begin{equation}
\label{eq:knn_loss}
\mathcal{L}_{k\text{-NN}}
=
\mathcal{L}_{k\text{-NN}}^{+}
+
\alpha_{\mathrm{neg}}
\mathcal{L}_{k\text{-NN}}^{-},
\end{equation}
where $\alpha_{\mathrm{neg}}$ controls the strength of the negative separation.

By jointly optimizing the RQ-VAE~\cite{ZeghidourLOST22}
quantization objective with the \texorpdfstring{$k$}{k}-NN pseudo-neighbor prefix consistency constraint, continuous code-neighborhood relations are translated into hierarchical prefix supervision in the discrete DocID space. 
The resulting globally static DocIDs are designed to organize code snippets into coarse-to-fine semantic regions while preserving fine-grained discriminability for generative retrieval.

\subsection{Dual-Phase DocID Generation Guidance}
\label{sec:dual_phase_docid_generation_guidance}

The globally fixed hierarchical DocID space constructed above provides a structured retrieval target, while effective retrieval further requires reliable query-conditioned navigation within this space.
To this end, \textsc{CodeHID} introduces \emph{Dual-Phase DocID Generation Guidance}, combining training-side ranking enhancement to discriminate confusable DocIDs with inference-side candidate constraint and prefix guidance to prioritize query-relevant generation paths.

\vspace{5pt}
\noindent\textbf{Training-Side Ranking Enhancement.}
The standard DocID generation objective mainly supervises the target identifier, without explicitly calibrating its preference against other legal candidates for the same query~\cite{Tay00NBM000GSCM22,WangHWMWCXCZL0022,0002DW23}.
To strengthen discrimination among confusable DocIDs, \textsc{CodeHID} augments query-to-DocID generation with local ranking supervision that identifies competitive candidates and preserves their relative relevance.

For a training query $q_i$, we first mine high-scoring non-target DocIDs from the valid DocID space:
\begin{equation}
\label{eq:hard_negative}
\mathcal{H}_i
=
\operatorname{TopK}_{d\in \mathcal{D}\setminus\{d_i^+\}}
R(q_i,d),
\end{equation}
where $d_i^+$ denotes the positive DocID, $\mathcal{D}$ is the valid DocID set, and $R(q_i,d)$ assigns a query-conditioned relevance score to candidate DocID $d$.
Specifically, $R(q_i,d)$ is the cosine similarity between the query representation and the representation of the code snippet indexed by DocID $d$. 

The resulting hard negatives are combined with the positive target to form the local candidate set:
\begin{equation}
\label{eq:dual_phase_candidates}
\mathcal{C}_i
=
\{d_i^+\}
\cup
\mathcal{H}_i .
\end{equation}
By selecting high-scoring alternatives rather than arbitrary negatives, $\mathcal{C}_i$ concentrates training on legal DocIDs that are more likely to compete with the target.

Hard-negative mining determines which DocIDs to contrast with the target, while the scores $R(q_i,d)$ additionally encode their relative competitiveness.
Treating $\mathcal{H}_i$ as an unordered negative set collapses the graded competition among prefix-sharing legal DocIDs into binary target--negative supervision.
We therefore preserve the soft ordering induced by $R(q_i,d)$ and transfer this local preference structure to the DocID generator through rank distillation.

Specifically, two score vectors are defined over the same local candidate set:
\begin{equation}
\label{eq:candidate_score}
\mathbf{r}_i^{a}
=
h_a(q_i,\mathcal{C}_i)
=
[r_i^{a,+},r_{i1}^{a,-},\dots,r_{iK}^{a,-}],
\quad
a\in\{\mathrm{S},\mathrm{T}\}.
\end{equation}
Here, $\mathrm{S}$ and $\mathrm{T}$ denote the student and teacher scoring sources, respectively.
The student scorer $h_{\mathrm{S}}=h_\theta$ is the \textsc{CodeHID} query-to-DocID generator and evaluates each candidate DocID by its teacher-forced conditional generation likelihood under $q_i$.
The teacher scorer $h_{\mathrm{T}}$ reuses the relevance function $R(q_i,d)$ in Eq.~\ref{eq:hard_negative} over the same candidate set.
Thus, $\mathbf{r}_i^{\mathrm{S}}$ represents the generator's current preference over $\mathcal{C}_i$, whereas $\mathbf{r}_i^{\mathrm{T}}$ retains the graded relevance already induced during candidate mining.

Since the two scoring mechanisms operate on different numerical scales, we align their relative preferences rather than their raw scores.
Both score vectors are converted into temperature-controlled local ranking distributions:
\begin{equation}
\label{eq:rank_distribution}
p_i^{a}
=
\operatorname{softmax}
\left(
\frac{\mathbf{r}_i^{a}}{\tau_{\mathrm{rank}}}
\right),
\quad
a\in\{\mathrm{S},\mathrm{T}\},
\end{equation}
where $\tau_{\mathrm{rank}}$ is the distillation temperature. 
The teacher distribution $p_i^{\mathrm{T}}$ captures the soft ordering among locally competitive DocIDs, while $p_i^{\mathrm{S}}$ reflects the corresponding preference of the generator.
The rank-distillation loss $\mathcal{L}_{\mathrm{rank}}$ transfers this local ordering structure to the student generator.

The final training objective combines the standard DocID generation objective with two complementary ranking signals:
\begin{equation}
\label{eq:dual_phase_train}
\left\{
\begin{aligned}
\mathcal{L}_{\mathrm{docid}}
&=
-\sum_{\ell=1}^{L}
\log
p_{\theta}
\left(
d_i^{(\ell)}
\mid
q_i,d_i^{(<\ell)}
\right),
\\
\mathcal{L}_{\mathrm{lex}}
&=
-\frac{1}{K}
\sum_{j=1}^{K}
\log
\sigma
\left(
r_i^{S,+}
-
r_{ij}^{S,-}
\right),
\\
\mathcal{L}_{\mathrm{rank}}
&=
\operatorname{KL}
\left(
p_i^{T}
\parallel
p_i^{S}
\right),
\\
\mathcal{L}_{\mathrm{guidance}}
&=
\mathcal{L}_{\mathrm{docid}}
+
\lambda_{\mathrm{lex}}\mathcal{L}_{\mathrm{lex}}
+
\lambda_{\mathrm{rank}}\mathcal{L}_{\mathrm{rank}} .
\end{aligned}
\right.
\end{equation}
$\mathcal{L}_{\mathrm{docid}}$ is the autoregressive cross-entropy objective over the target DocID tokens.
$\mathcal{L}_{\mathrm{lex}}$ explicitly encourages the generator to assign higher scores to the positive DocID than to mined hard negatives through pairwise preference learning.
$\mathcal{L}_{\mathrm{rank}}$ transfers the teacher-induced soft ranking distribution over locally competitive DocIDs to the student generator via KL divergence.
Coefficients $\lambda_{\mathrm{lex}}$ and $\lambda_{\mathrm{rank}}$ control the two auxiliary objectives.
Together, lexical consistency and rank distillation strengthen target discrimination while preserving finer-grained preference among confusable DocID paths.

\vspace{5pt}
\noindent\textbf{Decoding-Side Candidate Constraint and Reranking.}
During inference, \textsc{CodeHID} restricts autoregressive generation to query-relevant DocID paths and refines their priorities in beam search.

For an input query $q$, query-conditioned retrieval over the offline corpus index ranks code candidates using the cosine similarity score $R(q,d)$ defined in Eq.~\ref{eq:hard_negative} and selects the top-$B$ candidates, whose corresponding DocIDs form $\mathcal{D}_q=\{d_1,d_2,\dots,d_B\}$.
These DocIDs define a query-specific trie for constrained autoregressive generation, where each decoding step only allows valid prefix extensions.
The generator therefore performs autoregressive DocID generation within this query-conditioned address space, with complete paths scored by their accumulated generation probabilities.

The relevance scores associated with the retrieved candidates further provide query-level guidance for path selection.
By propagating these scores along their hierarchical DocID paths, \textsc{CodeHID} derives prefix-level scores that guide intermediate beam expansions.
At the $\ell$-th decoding step, the generation log-probability of an admissible token $v$ is adjusted as
\begin{equation}
\label{eq:dual_phase_logit}
\tilde{z}_{v}
=
z_{v}
+
\beta b_{\mathrm{lex}}
\left(
q,d^{(<\ell)},v
\right),
\end{equation}
where $z_v$ is the original autoregressive generation log-probability,
$b_{\mathrm{lex}}(q,d^{(<\ell)},v)$ denotes the query-conditioned prefix score for extending $d^{(<\ell)}$ with token $v$, and the lexical bonus weight $\beta$ controls the contribution of this guidance.
The resulting score integrates the generator's learned preference with query-specific evidence over the constrained DocID paths.

The guided token scores are accumulated along each complete DocID path:
\begin{equation}
\label{eq:dual_phase_decode_score}
\tilde{s}_{\theta}(d\mid q)
=
\sum_{\ell=1}^{L}
\tilde{z}_{d^{(\ell)}} .
\end{equation}
Here, $\tilde{z}_{d^{(\ell)}}$ denotes the guided score of the token $d^{(\ell)}$ at level $\ell$. The final prediction is obtained by
\begin{equation}
\label{eq:dual_phase_decode}
\hat{d}
=
\arg\max_{d\in\mathcal{D}_q^{\mathrm{trie}}}
\tilde{s}_{\theta}(d\mid q),
\end{equation}
where $\mathcal{D}_q^{\mathrm{trie}}$ denotes the valid prefix search space induced by the query-related candidate DocIDs.
The predicted DocID $\hat{d}$ is then resolved to its corresponding code snippet through the static DocID-to-code lookup.

Together with training-side ranking enhancement, this decoding procedure couples local candidate discrimination with coarse-to-fine path selection over the same hierarchical DocID space.

\section{Experiments}

\subsection{Experimental Setting}

\subsubsection{Datasets}

We use two code retrieval benchmarks: CoSQA and ProCQA. 
CoSQA~\cite{HuangTSG0J0D20} is a human-annotated dataset for Python code search and code question answering, while 
ProCQA~\cite{LiZYOR24} is a large-scale community-based programming question answering dataset constructed from StackOverflow. 

For CoSQA, we use its official train/dev/test split and retain only query--code pairs with positive annotations ($label=1$) as generation targets. 
ProCQA requires additional preprocessing due to its community-generated nature. 
We therefore construct a cleaning pipeline following the filtering criteria of CodeXGLUE AdvTest~\cite{LuGRHSBCDJTLZSZ21}, removing invalid, overly short, low-information, duplicated, and potentially contaminated code snippets. 
To suppress residual duplication caused by superficial syntactic variation, the retained code is further normalized through alpha-renaming of identifiers and literal standardization, and normalized AST fingerprints are used to detect structurally duplicate or near-duplicate code groups. 
When multiple valid candidate snippets are associated with the same question, we keep the highest-quality snippet as the target code, which yields more reliable query-code pairs and reduces the influence of noisy answers on training and evaluation. 
After cleaning, the candidate pools contain approximately 406.5K Python samples and 194.6K Java samples.
Following the CoSQA train/dev/test setting, we randomly sample about 30K instances for each language and construct an 8:2:2 train/dev/test split for ProCQA.

\subsubsection{Evaluation Metrics}

We evaluate retrieval effectiveness using standard measures Hit@1, Hit@3, Hit@5, and MRR@20.

\subsubsection{Baselines}
We use four representative baselines:
\begin{itemize}[leftmargin=*,topsep=2pt]
    \item \textbf{Lexical sparse retrieval.}
    BM25~\cite{RobertsonWJHG94} is adopted as a classical sparse retrieval baseline, which relies on lexical term matching between natural language queries and code snippets.

    \item \textbf{Pre-trained code models.}
    CodeBERT~\cite{FengGTDFGS0LJZ20} and UniXcoder~\cite{GuoLDW0022} are selected as representative pre-trained code retrieval models. 
    They learn semantic representations for natural language and code. Code candidates with the highest similarity to the query are returned as results.

    \item \textbf{Dense code retrieval models.}
    CodeSage~\cite{ZhangATDNR0X24}, OASIS~\cite{GaoZLYZCZ025}, and CodeXEmbed~\cite{LiuMJSXZY25} are used as representative semantic code retrieval baselines.

    \item \textbf{Generative retrieval models.}
    DSI~\cite{Tay00NBM000GSCM22}, NCI~\cite{WangHWMWCXCZL0022}, GLEN~\cite{00010L23}, and RIPOR~\cite{Zeng0JSWZ24} are used as representative generative retrieval baselines. 
    DSI and NCI establish the query-to-identifier generation paradigm, while GLEN and RIPOR further improve identifier learning and ranking optimization. 

\end{itemize}

\subsubsection{Implementation Details.}
We implement \textsc{CodeHID} in two stages.
For offline DocID construction, Pseudo-Neighbor Guided DocID Learning assigns each code snippet a globally static four-level DocID with a codebook size $M=256$.
The pseudo-neighbor prefix supervision uses neighborhood size $k=8$, prefix-consistency loss weight $\lambda_{k\text{-NN}}=0.03$, negative separation weight $\alpha_{\mathrm{neg}}=0.6$, RQ-VAE temperature $\gamma=0.35$, and level-wise weights $w_\ell=[0.8,0.35,0.12,0.03]$, where the level-wise similarity thresholds $\tau_\ell$ are automatically determined using quantiles $[0.85,0.95,0.99,0.999]$.
For query-to-DocID generation, we use a frozen GraphCodeBERT~\cite{GuoRLFT0ZDSFTDC21} encoder and a Qwen2.5-Coder-7B decoder fine-tuned with LoRA adapters.
The generator is fine-tuned for 12 epochs with AdamW, using a learning rate of $5\times10^{-5}$, weight decay of $0.01$, and a random seed of 42.
The training-side ranking enhancement uses $K=4$ hard negatives, $\lambda_{\mathrm{lex}}=0.06$, $\lambda_{\mathrm{rank}}=0.03$, and $\tau_{\mathrm{rank}}=1.0$.
During inference, Dual-Phase DocID Generation Guidance performs constrained beam search over valid DocID prefixes with beam size 20, and applies prefix-level reranking with candidate set size $B=3000$ and lexical bonus coefficient $\beta=0.10$.
All inference-time signals are computed from the input query and the retrieval corpus, without using gold relevance labels.

\subsection{Experimental Results}
\subsubsection{Overall Performance}
\label{sec:overall_performance}

Tab.~\ref{tab:overall_performance} reports the overall performance. 
Overall, \textsc{CodeHID} consistently presents the strongest overall performance across all datasets. 

\vspace{5pt}
\noindent\textbf{Comparison with matching-based retrieval.}  
Compared with matching-based methods (sparse retrieval, pre-trained code models, and dense code retrieval models), \textsc{CodeHID} shows better results. 
On CoSQA, \textsc{CodeHID} achieves the best Hit@1, Hit@3, and MRR@20, while remaining competitive on Hit@5. On ProCQA, \textsc{CodeHID} outperforms baselines in all cases. 
This indicates that its early-rank improvements do not come at the cost of candidate coverage; instead, \textsc{CodeHID} improves top-ranked ordering while preserving comparable top-$k$ coverage.

These results reveal a bottleneck beyond similarity scoring: when multiple candidates are plausible, pairwise matching alone may fail to distinguish them effectively.
\textsc{CodeHID} addresses this issue by organizing code snippets into a globally static hierarchical DocID space and retrieving them through hierarchical address generation.
Thus, its advantage comes from structuring the retrieval target space rather than merely replacing the matching function.

\begin{table*}[t]
\centering
\setlength{\tabcolsep}{4.2pt}
\caption{Overall performance comparison. The best result for each metric is highlighted in bold.}
\label{tab:overall_performance}
\vspace{-10pt}
\scalebox{0.95}{
\begin{tabular}{lcccc|cccc|cccc}
\toprule
\multirow{2}{*}{Model}
& \multicolumn{4}{c|}{CoSQA}
& \multicolumn{4}{c|}{ProCQA-Python}
& \multicolumn{4}{c}{ProCQA-Java} \\
\cmidrule(lr){2-5}
\cmidrule(lr){6-9}
\cmidrule(lr){10-13}
& Hit@1 & Hit@3 & Hit@5 & MRR@20
& Hit@1 & Hit@3 & Hit@5 & MRR@20
& Hit@1 & Hit@3 & Hit@5 & MRR@20 \\
\midrule
BM25
& 0.252 & 0.408 & 0.486 & 0.358
& 0.316 & 0.386 & 0.418 & 0.363
& 0.360 & 0.433 & 0.471 & 0.411 \\

\midrule

CodeBERT
& 0.284 & 0.514 & 0.633 & 0.440
& 0.172 & 0.243 & 0.275 & 0.221
& 0.181 & 0.257 & 0.299 & 0.236 \\

UniXcoder
& 0.518 & 0.741 & \textbf{0.850} & 0.654
& 0.362 & 0.452 & 0.491 & 0.423
& 0.363 & 0.462 & 0.504 & 0.429 \\

\midrule

CodeSage
& 0.444 & 0.646 & 0.732 & 0.576
& 0.248 & 0.330 & 0.367 & 0.308
& 0.266 & 0.364 & 0.403 & 0.334 \\

OASIS
& 0.435 & 0.674 & 0.767 & 0.583
& 0.342 & 0.436 & 0.481 & 0.406
& 0.347 & 0.452 & 0.500 & 0.417 \\

CodeXEmbed
& 0.470 & 0.700 & 0.796 & 0.611
& 0.208 & 0.291 & 0.327 & 0.264
& 0.216 & 0.304 & 0.347 & 0.277 \\

\midrule
DSI
& 0.319 & 0.417 & 0.440 & 0.375
& 0.056 & 0.166 & 0.326 & 0.227
& 0.102 & 0.315 & 0.509 & 0.297 \\

NCI
& 0.498 & 0.633 & 0.709 & 0.587
& 0.406 & 0.622 & 0.731 & 0.553
& 0.325 & 0.573 & 0.694 & 0.491 \\

GLEN
& 0.380 & 0.569 & 0.655 & 0.500
& 0.262 & 0.433 & 0.534 & 0.391
& 0.318 & 0.534 & 0.653 & 0.464 \\

RIPOR
& 0.256 & 0.412 & 0.489 & 0.361
& 0.184 & 0.307 & 0.397 & 0.307
& 0.200 & 0.357 & 0.471 & 0.342 \\

\midrule
\textbf{\textsc{CodeHID}}
& \textbf{0.744} & \textbf{0.783} & 0.792 & \textbf{0.766}
& \textbf{0.597} & \textbf{0.773} & \textbf{0.828} & \textbf{0.706}
& \textbf{0.520} & \textbf{0.734} & \textbf{0.825} & \textbf{0.654} \\
\bottomrule
\end{tabular}
}
\end{table*}

\begin{table}[t]
\centering
\caption{Comparison at comparable model scales.}
\vspace{-10pt}
\label{tab:model_scale}
\setlength{\tabcolsep}{3pt}
\resizebox{\columnwidth}{!}{
\begin{tabular}{lcccccc}
\toprule
\multirow{2}{*}{Model} &
\multicolumn{2}{c}{CoSQA} &
\multicolumn{2}{c}{ProCQA-Python} &
\multicolumn{2}{c}{ProCQA-Java} \\
\cmidrule(lr){2-3}
\cmidrule(lr){4-5}
\cmidrule(lr){6-7}
& Hit@1 & MRR@20
& Hit@1 & MRR@20
& Hit@1 & MRR@20 \\
\midrule
CodeSage$_{\text{1.3B}}$
& 0.444 & 0.576
& 0.248 & 0.308
& 0.266 & 0.334 \\

OASIS$_{\text{1.5B}}$
& 0.435 & 0.583
& 0.342 & 0.406
& 0.347 & 0.417 \\

CodeXEmbed$_{\text{2B}}$
& 0.470 & 0.611
& 0.208 & 0.264
& 0.216 & 0.277 \\

\midrule
\textsc{CodeHID}$_{\text{1.5B}}$
& \textbf{0.565} & \textbf{0.614}
& \textbf{0.447} & \textbf{0.606}
& \textbf{0.399} & \textbf{0.565} \\
\bottomrule
\end{tabular}
}
\end{table}

\vspace{5pt}
\noindent\textbf{Comparison with generative retrieval.} 
Although DSI, NCI, GLEN, and RIPOR formulate retrieval as DocID generation, they do not show consistent advantages for code retrieval.
NCI achieves competitive top-$k$ coverage on ProCQA, but its Hit@1 and MRR@20 remain below \textsc{CodeHID}, indicating that generatable DocIDs alone cannot reliably prioritize the correct decoding path. 
Likewise, GLEN's lexical index learning and RIPOR's prefix-oriented optimization do not consistently improve performance.
These results suggest that local refinements to generic generative retrieval are insufficient to solve the identifier-space organization problem.

\textsc{CodeHID} instead jointly models how the target space is constructed and how it is accessed.
Pseudo-Neighbor Guided DocID Learning organizes DocID prefixes according to code semantics, while Dual-Phase DocID Generation Guidance improves query-conditioned path selection.
Its advantage therefore stems from coupling a semantically organized hierarchical DocID space with reliable query-conditioned generation.

\vspace{5pt}
\noindent\textbf{Comparison at comparable model scales.} 
In the main experiments, we report \textsc{CodeHID} using Qwen2.5-Coder-7B. 
To investigate whether \textsc{CodeHID}'s improvements come from the larger model size, we further compare a smaller \textsc{CodeHID} variant (using Qwen2.5-Coder-1.5B) with parameter-comparable baselines in Tab.~\ref{tab:model_scale}, where model size is shown as a subscript.
We can see that \textsc{CodeHID}$_{\text{1.5B}}$ maintains stronger performance compared to baselines with 1.3B-2B parameters, suggesting that the gains of \textsc{CodeHID} come from its hierarchical DocID organization and query-conditioned generation mechanism, instead of the larger model size.

\subsubsection{Ablation Study}
We report the ablation study on CoSQA dataset in Tab.~\ref{tab:ablation}.
The complete model achieves the best results, indicating that Pseudo-Neighbor Guided DocID Learning and Dual-Phase DocID Generation Guidance jointly support the overall improvements.

\vspace{5pt}
\noindent\textbf{Structured DocID space enables effective path guidance.}
The main ablations show that plain quantization-based
DocID generation is insufficient for effective code retrieval.
As shown in Tab.~\ref{tab:ablation}, removing pseudo-supervised DocID learning reduces Hit@1 from 0.744 to 0.399 despite query-conditioned guidance, indicating that without stable semantic partitions in DocID prefixes, training-side ranking and inference-side prefix reranking cannot reliably distinguish competing candidates through coarse-to-fine generation.
The primary benefit of \textsc{CodeHID}, therefore, lies in translating continuous code-neighborhood semantics into a hierarchical DocID space that is more amenable to generation.

\vspace{5pt}
\noindent\textbf{Hierarchical pseudo labels shape the generation structure.} 
The pseudo-label variants show that \texorpdfstring{$k$}{k}-NN pseudo-neighbor supervision regulates prefix sharing at different depths rather than merely grouping similar code snippets.
Removing coarse- or fine-level pseudo labels reduces
Hit@1 to 0.364 and 0.390, respectively, reflecting their
complementary roles: coarse-level labels promote stable
semantic partitioning in early generation steps and narrow
the candidate space, while fine-level labels preserve
target-level distinctions among functionally similar but
implementation-distinct snippets at deeper prefix levels.
In this sense, pseudo-supervised DocID learning provides a more ordered and discriminative target structure for autoregressive generation.

\vspace{5pt}
\noindent\textbf{Dual-phase query guidance selects the target path.} 
The query-conditioned guidance variants indicate that, even with a fixed DocID space, the generation process still requires query signals to determine the correct path.
During training, ranking distillation and lexical consistency strengthen the model's preference for the positive DocID over hard negatives.
During inference, Prefix Scoring guides intermediate
prefix expansion, while Query Scoring evaluates complete
DocIDs. 
The ablations reveal a non-monotonic interaction:
retaining either signal alone yields lower Hit@1
(0.610/0.607) than disabling both (0.728), suggesting
that one-sided guidance may misalign prefix exploration
with final candidate relevance. 
Thus, valid-prefix constraints ensure that generated paths
map to existing code, while query-conditioned guidance
selects the most query-relevant path among valid alternatives.

\begin{table}[t]
\centering
\caption{Ablation study of \textsc{CodeHID} on CoSQA. }
\label{tab:ablation}
\vspace{-10pt}
\scriptsize
\setlength{\tabcolsep}{3.0pt}
\renewcommand{\arraystretch}{0.88}
\resizebox{\columnwidth}{!}{
\begin{tabular}{lcccc}
\toprule
\textbf{Setting} & \textbf{H@1} & \textbf{H@3} & \textbf{H@5} & \textbf{MRR@20} \\
\midrule

\multicolumn{5}{l}{\textit{Main ablations}} \\
w/o DocID Learning + Query Signals
& 0.259 & 0.355 & 0.412 & 0.333 \\
w/o DocID Learning
& 0.399 & 0.504 & 0.536 & 0.462 \\
w/o Train-time Query Signals
& 0.355 & 0.441 & 0.479 & 0.414 \\
w/o Infer-time Query Signals
& 0.728 & 0.764 & 0.783 & 0.749 \\
\textbf{\textsc{CodeHID} (Full)}
& \textbf{0.744} & \textbf{0.783} & \textbf{0.792} & \textbf{0.766} \\

\midrule
\multicolumn{5}{l}{\textit{Pseudo-label DocID variants}} \\
w/o Coarse Pseudo Labels
& 0.364 & 0.457 & 0.511 & 0.427 \\
w/o Fine Pseudo Labels
& 0.390 & 0.470 & 0.505 & 0.444 \\

\midrule
\multicolumn{5}{l}{\textit{Query-conditioned generation variants}} \\
w/o Rank Distillation
& 0.361 & 0.444 & 0.482 & 0.418 \\
w/o Lexical Consistency
& 0.351 & 0.447 & 0.482 & 0.413 \\
w/o Prefix Scoring
& 0.610 & 0.634 & 0.642 & 0.623 \\
w/o Query Scoring
& 0.607 & 0.629 & 0.642 & 0.623 \\

\bottomrule
\end{tabular}
}

\vspace{2pt}
\footnotesize
\raggedright
\textit{Note:} H@\(k\) denotes Hit@\(k\).
\vspace{-5pt}
\end{table}

\subsubsection{Robustness to Hyperparameters}
We further examine the robustness of \textsc{CodeHID} w.r.t. key hyperparameters on CoSQA.

\begin{figure}[t]
    \centering
    \includegraphics[width=1\columnwidth]
    {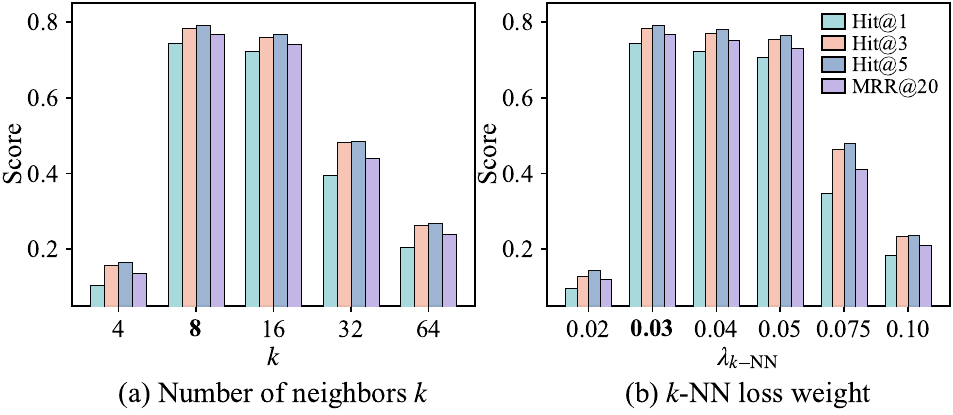}
    \caption{
    Sensitivity analysis of pseudo-neighbor supervision on CoSQA.
    (a) varies the number of pseudo neighbors $k$ with $\lambda_{k\text{-NN}}=0.03$ fixed, while
    (b) varies the prefix-consistency loss weight $\lambda_{k\text{-NN}}$ with $k=8$ fixed.
    The best overall performance is observed at $k=8$ and $\lambda_{k\text{-NN}}=0.03$, suggesting that moderate pseudo-neighbor supervision provides a more effective prefix organization for DocID generation.
    }
    \label{fig:rqvae_knn_sensitivity}
\end{figure}

\vspace{5pt}
\noindent\textbf{Sensitivity to neighborhood supervision.}  
Fig.~\ref{fig:rqvae_knn_sensitivity} examines two key
hyperparameters: neighborhood size
\texorpdfstring{$k$}{k} and prefix-consistency loss weight
\texorpdfstring{$\lambda_{k\text{-NN}}$}{lambda k-NN},
which control the coverage and semantic precision of
pseudo-neighbor relations and the strength of their
supervision in quantization learning, respectively.

\vspace{5pt}
\noindent\textbf{a) Effect of neighborhood coverage.} 
\textsc{CodeHID} achieves the best performance at
\texorpdfstring{$k=8$}{k=8} and remains close at
\texorpdfstring{$k=16$}{k=16}, favoring moderate
neighborhood coverage. When \texorpdfstring{$k$}{k} is
too small, unstable neighborhood signals make the
quantization behave closer to ordinary quantization,
leaving DocID prefixes with limited retrieval semantics.
Conversely, overly large neighborhoods introduce weakly
related samples, reducing the precision of level-wise
pseudo labels and causing target-distinct snippets to
share overly similar prefixes, thereby weakening
deep-level discrimination.

\vspace{5pt}
\noindent\textbf{b) Effect of prefix-consistency strength.} 
The model performs best at \texorpdfstring{$\lambda_{k\text{-NN}}=0.03$}{lambda k-NN=0.03} and remains stable around 0.04--0.05, indicating that prefix consistency should complement rather than dominate the quantization objective. 
An overly weak constraint fails to translate continuous semantic neighborhoods into discrete prefix structures, whereas an overly strong one overfits imperfect \texorpdfstring{$k$}{k}-NN relations, disrupting reconstruction and residual quantization and compromising target-level separability.

\vspace{5pt} 
Overall, the stable behavior under moderate \texorpdfstring{$k$}{k} and \texorpdfstring{$\lambda_{k\text{-NN}}$}{lambda k-NN} shows that pseudo-supervised DocID learning relies on a balance between structural constraint and quantization learning.
Neighborhood supervision must be strong enough to shape hierarchical prefixes, yet precise enough to avoid compressing weakly related code snippets into overly similar generation paths.

\begin{figure}[t]
    \centering
    \setlength{\abovecaptionskip}{2pt}
    \includegraphics[
        width=0.88\columnwidth,
    ]{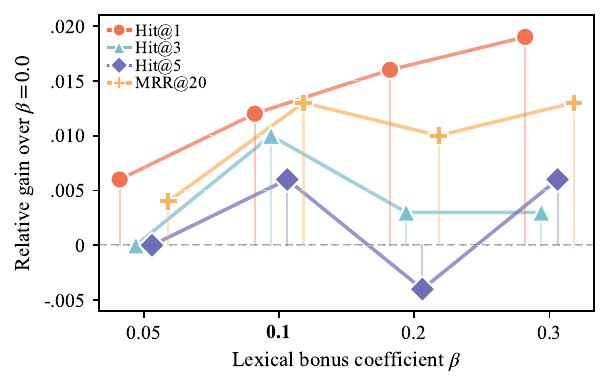}
    \caption{
    Sensitivity analysis of decoding-side lexical guidance by varying the lexical bonus coefficient $\beta$ used in prefix-level reranking on CoSQA.
    The y-axis reports the relative gain over constrained decoding without a lexical bonus, i.e., $\beta=0.0$.
    The bold tick mark denotes the setting used in the main experiments, $\beta=0.1$. 
    }
    \label{fig:lexical_guidance_sensitivity}
\end{figure}

\vspace{5pt}
\noindent\textbf{Sensitivity to decoding-side lexical guidance.} 
Fig.~\ref{fig:lexical_guidance_sensitivity} examines the
effect of lexical bonus coefficient $\beta$ in prefix-level
reranking, which balances autoregressive DocID generation
probabilities with query-specific lexical evidence.
Moderate lexical guidance improves early-rank metrics,
particularly Hit@1 and MRR@20, by calibrating the priorities of query-relevant prefixes within the constrained decoding space. 
At $\beta=0.1$, used in the main experiments, lexical
evidence adjusts prefix priorities without overriding the
learned generation distribution. 
However, stronger guidance may overemphasize surface lexical overlap, potentially improving Hit@1 while perturbing the top-$k$ candidate distribution and causing fluctuations in Hit@3 and Hit@5.
Thus, lexical guidance serves as a local path-calibration
signal rather than an alternative scoring objective.

\subsubsection{In-depth Analysis}

This section further examines how the design of \textsc{CodeHID} translates into more reliable rank-one retrieval.
The analysis focuses on three aspects: how the pseudo-supervised hierarchical DocID space reduces candidate ambiguity along the generation path, how Dual-Phase DocID Generation Guidance steers decoding toward legal paths that better match the query intent, and whether \textsc{CodeHID} can make more precise first-rank decisions when the target code has already been recalled among the candidates.

\vspace{5pt}
\noindent\textbf{Generation-friendly DocID target space.}
\textsc{CodeHID} translates continuous code-neighborhood
relations into hierarchical prefix-sharing structures
through \texorpdfstring{$k$}{k}-NN pseudo-neighbor relations and level-wise pseudo labels. 
Shallow prefixes capture coarse semantic regions to narrow the candidate space, while deeper prefixes distinguish implementation variants among functionally similar snippets, forming a coarse-to-fine structure for autoregressive generation.

This perspective also clarifies why the learned target
space matters for rank-one retrieval: without meaningful
prefix organization, target codes may remain entangled
with similar candidates along generation paths, leaving
query-conditioned guidance to operate over a weakly
structured search space. 
The ablation results support this interpretation: removing pseudo-supervised DocID learning weakens the model despite retaining query guidance, while removing coarse or fine pseudo labels impairs early semantic partitioning or deep-level target discrimination, respectively. 
These results suggest that \textsc{CodeHID}'s rank-one advantage is grounded in a generation-friendly DocID space, where meaningful prefix organization enables query-conditioned guidance to reliably select the target path among legal alternatives.

\vspace{5pt}
\noindent\textbf{Query-conditioned path prioritization.}
The query-conditioned variants reveal that the main challenge in a fixed DocID space is not merely path validity, but the relative prioritization of legal paths under the query.
For functionally similar or implementation-related snippets, raw autoregressive probabilities may favor plausible but non-target paths, suppressing the target during generation.
Dual-Phase DocID Generation Guidance mitigates this priority mismatch by combining training-side ranking preferences with inference-side prefix reranking, keeping the target competitive against similar legal alternatives.

This interpretation also explains the degradation after removing Query Scoring or Prefix Scoring: even when generated DocIDs remain valid, beam search may favor structurally valid and locally plausible paths that do not match the query intent.
The value of dual-phase guidance therefore lies not in
enlarging the search space, but in recalibrating path
priorities within the fixed legal DocID space for more
reliable rank-one discrimination.

\begin{table}[t]
\centering
\caption{\textsc{CodeHID}'s rank-one discrimination capability on baselines' top-5 failures.}
\label{tab:rank1_promotion}
\vspace{-10pt}
\resizebox{\columnwidth}{!}{
\setlength{\tabcolsep}{4pt}
\begin{tabular}{lrrrr}
\toprule
Baseline & \(N^b_{5\setminus1}\) & \textsc{CodeHID}@1 & \textsc{CodeHID}@5 & Recovery@1 \\
\midrule
BM25       & 73  & 55  & 58  & 75.3\% \\
CodeBERT   & 109 & 81  & 86  & 74.3\% \\
UniXcoder  & 104 & 88  & 90  & 84.6\% \\
OASIS      & 104 & 83  & 87  & 79.8\% \\
CodeXEmbed & 111 & 91  & 95  & 82.0\% \\
\midrule
\textbf{All} & \textbf{249} & \textbf{190} & \textbf{201} & \textbf{76.3\%} \\
\bottomrule
\end{tabular}
}
\end{table}

\vspace{5pt}
\noindent\textbf{Rank-one discrimination capability.} 
We further analyze \textsc{CodeHID}'s rank-one discrimination capability.
For each baseline \(b\), we define \(S_b\) as the set of queries where the target code is retrieved within the top-5 but not at rank one.
The size of this set, \(N^b_{5\setminus1}\), represents baseline top-5 failures.
\textsc{CodeHID} independently performs generative retrieval on queries in \(S_b\) without accessing the baseline-retrieved candidates.
\textsc{CodeHID}@\(K\) denotes the number of queries in \(S_b\) where \textsc{CodeHID} retrieves the target code within the top-\(K\) positions.
Recovery@1 ($=$\textsc{CodeHID}$@1/N^b_{5\setminus1}$) measures how often \textsc{CodeHID} places the target code at rank one for queries in \(S_b\).

Tab.~\ref{tab:rank1_promotion} shows that \textsc{CodeHID} achieves consistently high Recovery@1 across \(S_b\) sets, indicating the robustness of \textsc{CodeHID} to retrieval failure patterns encountered by baselines (i.e., baselines fail to discriminate targets and plausible candidates but \textsc{CodeHID} succeeds).
\textsc{CodeHID}'s ability to recover these failures indicates that its improvement comes from more accurate target selection within a structured retrieval space.
The pseudo-supervised hierarchical DocID space reduces ambiguity among competing generation paths, while Dual-Phase DocID Generation Guidance improves query-conditioned path selection within the valid DocID space.
Together, these mechanisms enable \textsc{CodeHID}'s stronger rank-one discrimination capability.
\section{Related Work}

\noindent\textbf{Code Retrieval.}
Code retrieval is commonly formulated as query-code matching.
For instance, CodeBERT~\cite{FengGTDFGS0LJZ20} and CodeRetriever~\cite{LiGSQZYQJCD22} learn shared semantic spaces for natural language and source code.
Structure-aware models further exploit program signals, including data flow in GraphCodeBERT~\cite{GuoRLFT0ZDSFTDC21}, identifier-aware modeling in CodeT5~\cite{0034WJH21}, and cross-modal pre-training in UniXcoder~\cite{GuoLDW0022}.
Although these methods improve code representations, retrieval is still typically performed over a flat candidate space where snippets are independently scored.
In contrast, \textsc{CodeHID} treats the organization of the retrieval target space as a central problem and addresses code snippets through hierarchical identifiers.

\vspace{5pt}
\noindent\textbf{Generative Retrieval and DocID Learning.}
Generative retrieval replaces candidate scoring with identifier generation.
DSI~\cite{Tay00NBM000GSCM22} introduces query-to-DocID generation, followed by studies on discrete auto-encoding~\cite{0001YCWZRCYRR23}, query-enriched descriptions~\cite{Tang0GCZWYC23}, and end-to-end indexing~\cite{YangSZHDSZ23}.
Multi-level identifier design has also been explored through residual quantization and retrieval-oriented hierarchical IDs: RQ-VAE~\cite{LeeKKCH22,ZeghidourLOST22} represents continuous inputs as stacked discrete codes, 
RQ-MoE~\cite{RQ-MoE} introduces input-dependent codebooks for residual quantization, 
TIGER~\cite{RajputMSKVHHT0S23} constructs semantic IDs for generative recommendation, and MERGE~\cite{ZhangLJZZLZLZ25} learns hierarchical identifiers with multi-level relevance signals.
Different from these studies, \textsc{CodeHID} studies DocID learning for code retrieval, where identifiers must capture both coarse functional similarity and fine-grained implementation-level discrimination~\cite{Tang0GR0C24}.

\vspace{5pt}
\noindent\textbf{Guided DocID Generation.}
Reliable DocID generation has been studied from the perspectives of decoding and path selection.
Autoregressive generation may suffer from early branching errors, where an incorrect prefix selection or premature pruning makes the target difficult to recover~\cite{0003R0YZCRR25}, and this issue becomes more severe as the corpus scales~\cite{Pradeep0GLZLM023}.
PAG~\cite{Zeng0Z24} mitigates this issue by using look-ahead signals from simultaneous decoding to guide autoregressive DocID generation.
These studies mainly improve decoding reliability once an identifier space is given, whereas code retrieval further requires distinguishing multiple valid paths for functionally similar but implementation-distinct snippets.
Dual-Phase DocID Generation Guidance addresses this ambiguity through training-side discrimination and inference-side prefix guidance.
\section{Conclusion}
This paper presents \textsc{CodeHID}, a generative code
retrieval framework that reformulates retrieval as
hierarchical semantic address generation over a structured
DocID space rather than flat candidate ranking.
It combines Pseudo-Neighbor Guided DocID Learning to
construct a semantically organized and discriminative
hierarchical index with Dual-Phase DocID Generation
Guidance to improve query-conditioned path selection.
Experiments across multiple benchmarks demonstrate its
effectiveness, highlighting the potential of hierarchical
semantic addressing for code retrieval and the importance
of structured target spaces and reliable query-conditioned
navigation.
Future work may explore code-structure-aware quantization,
incorporating relations such as call and inheritance
dependencies, and extend \textsc{CodeHID} to evolving
code corpora through continual learning.

\begin{acks}
This work is supported by the New Generation Artificial Intelligence-National Science and Technology Major Project (No.~2025ZD0122704), the National Natural Science Foundation of China (No.~62572410, No.~62602536), the Natural Science Foundation of Xiamen, China (No.~3502Z202471028, No.~3502Z202571028), and the Fundamental Research Funds for the Central Universities, Xiamen University (No.~20720250171).
\end{acks}

\clearpage

\section*{Ethical Considerations}
This work focuses on generative code retrieval by organizing code snippets with hierarchical DocIDs over existing benchmark datasets. The proposed method does not execute retrieved code, make decisions about individuals, or introduce additional collection of personal or sensitive information. 
Therefore, we do not identify significant ethical concerns associated with this work.

\bibliographystyle{ACM-Reference-Format}
\bibliography{ref}

@inproceedings{FengGTDFGS0LJZ20,
  author       = {Zhangyin Feng and
                  Daya Guo and
                  Duyu Tang and
                  Nan Duan and
                  Xiaocheng Feng and
                  Ming Gong and
                  Linjun Shou and
                  Bing Qin and
                  Ting Liu and
                  Daxin Jiang and
                  Ming Zhou},
  title        = {CodeBERT: {A} Pre-Trained Model for Programming and Natural Languages},
  booktitle    = {{EMNLP} (Findings)},
  pages        = {1536--1547},
  year         = {2020}
}

@inproceedings{GuoRLFT0ZDSFTDC21,
  author       = {Daya Guo and
                  Shuo Ren and
                  Shuai Lu and
                  Zhangyin Feng and
                  Duyu Tang and
                  Shujie Liu and
                  Long Zhou and
                  Nan Duan and
                  Alexey Svyatkovskiy and
                  Shengyu Fu and
                  Michele Tufano and
                  Shao Kun Deng and
                  Colin B. Clement and
                  Dawn Drain and
                  Neel Sundaresan and
                  Jian Yin and
                  Daxin Jiang and
                  Ming Zhou},
  title        = {GraphCodeBERT: Pre-training Code Representations with Data Flow},
  booktitle    = {{ICLR}},
  year         = {2021}
}

@inproceedings{GuoLDW0022,
  author       = {Daya Guo and
                  Shuai Lu and
                  Nan Duan and
                  Yanlin Wang and
                  Ming Zhou and
                  Jian Yin},
  title        = {UniXcoder: Unified Cross-Modal Pre-training for Code Representation},
  booktitle    = {{ACL} },
  pages        = {7212--7225},
  year         = {2022}
}

@inproceedings{Tay00NBM000GSCM22,
  author       = {Yi Tay and
                  Vinh Tran and
                  Mostafa Dehghani and
                  Jianmo Ni and
                  Dara Bahri and
                  Harsh Mehta and
                  Zhen Qin and
                  Kai Hui and
                  Zhe Zhao and
                  Jai Prakash Gupta and
                  Tal Schuster and
                  William W. Cohen and
                  Donald Metzler},
  title        = {Transformer Memory as a Differentiable Search Index},
  booktitle    = {NeurIPS},
  year         = {2022}
}

@inproceedings{ZhangLJZZLZLZ25,
  author       = {Fuwei Zhang and
                  Xiaoyu Liu and
                  Xinyu Jia and
                  Yingfei Zhang and
                  Shuai Zhang and
                  Xiang Li and
                  Fuzhen Zhuang and
                  Wei Lin and
                  Zhao Zhang},
  title        = {Multi-level Relevance Document Identifier Learning for Generative
                  Retrieval},
  booktitle    = {{ACL}},
  pages        = {10066--10080},
  year         = {2025}
}

@article{ZeghidourLOST22,
  author       = {Neil Zeghidour and
                  Alejandro Luebs and
                  Ahmed Omran and
                  Jan Skoglund and
                  Marco Tagliasacchi},
  title        = {SoundStream: An End-to-End Neural Audio Codec},
  journal      = {{IEEE} {ACM} Trans. Audio Speech Lang. Process.},
  volume       = {30},
  pages        = {495--507},
  year         = {2022}
}

@inproceedings{Zeng0Z24,
  author       = {Hansi Zeng and
                  Chen Luo and
                  Hamed Zamani},
  title        = {Planning Ahead in Generative Retrieval: Guiding Autoregressive Generation
                  through Simultaneous Decoding},
  booktitle    = {{SIGIR}},
  pages        = {469--480},
  year         = {2024}
}

@inproceedings{RajputMSKVHHT0S23,
  author       = {Shashank Rajput and
                  Nikhil Mehta and
                  Anima Singh and
                  Raghunandan Hulikal Keshavan and
                  Trung Vu and
                  Lukasz Heldt and
                  Lichan Hong and
                  Yi Tay and
                  Vinh Q. Tran and
                  Jonah Samost and
                  Maciej Kula and
                  Ed H. Chi and
                  Mahesh Sathiamoorthy},
  title        = {Recommender Systems with Generative Retrieval},
  booktitle    = {NeurIPS},
  year         = {2023}
}

@inproceedings{LiGSQZYQJCD22,
  author       = {Xiaonan Li and
                  Yeyun Gong and
                  Yelong Shen and
                  Xipeng Qiu and
                  Hang Zhang and
                  Bolun Yao and
                  Weizhen Qi and
                  Daxin Jiang and
                  Weizhu Chen and
                  Nan Duan},
  title        = {CodeRetriever: {A} Large Scale Contrastive Pre-Training Method for
                  Code Search},
  booktitle    = {{EMNLP}},
  pages        = {2898--2910},
  year         = {2022}
}

@inproceedings{0034WJH21,
  author       = {Yue Wang and
                  Weishi Wang and
                  Shafiq R. Joty and
                  Steven C. H. Hoi},
  title        = {CodeT5: Identifier-aware Unified Pre-trained Encoder-Decoder Models
                  for Code Understanding and Generation},
  booktitle    = {{EMNLP}},
  pages        = {8696--8708},
  year         = {2021}
}

@inproceedings{WangHWMWCXCZL0022,
  author       = {Yujing Wang and
                  Yingyan Hou and
                  Haonan Wang and
                  Ziming Miao and
                  Shibin Wu and
                  Qi Chen and
                  Yuqing Xia and
                  Chengmin Chi and
                  Guoshuai Zhao and
                  Zheng Liu and
                  Xing Xie and
                  Hao Sun and
                  Weiwei Deng and
                  Qi Zhang and
                  Mao Yang},
  title        = {A Neural Corpus Indexer for Document Retrieval},
  booktitle    = {NeurIPS},
  year         = {2022}
}

@inproceedings{00010L23,
  author       = {Sunkyung Lee and
                  Minjin Choi and
                  Jongwuk Lee},
  title        = {{GLEN:} Generative Retrieval via Lexical Index Learning},
  booktitle    = {{EMNLP}},
  pages        = {7693--7704},
  year         = {2023}
}

@inproceedings{0001YCWZRCYRR23,
  author       = {Weiwei Sun and
                  Lingyong Yan and
                  Zheng Chen and
                  Shuaiqiang Wang and
                  Haichao Zhu and
                  Pengjie Ren and
                  Zhumin Chen and
                  Dawei Yin and
                  Maarten de Rijke and
                  Zhaochun Ren},
  title        = {Learning to Tokenize for Generative Retrieval},
  booktitle    = {NeurIPS},
  year         = {2023}
}

@inproceedings{Tang0GCZWYC23,
  author       = {Yubao Tang and
                  Ruqing Zhang and
                  Jiafeng Guo and
                  Jiangui Chen and
                  Zuowei Zhu and
                  Shuaiqiang Wang and
                  Dawei Yin and
                  Xueqi Cheng},
  title        = {Semantic-Enhanced Differentiable Search Index Inspired by Learning
                  Strategies},
  booktitle    = {{KDD}},
  pages        = {4904--4913},
  year         = {2023}
}

@inproceedings{YangSZHDSZ23,
  author       = {Tianchi Yang and
                  Minghui Song and
                  Zihan Zhang and
                  Haizhen Huang and
                  Weiwei Deng and
                  Feng Sun and
                  Qi Zhang},
  title        = {Auto Search Indexer for End-to-End Document Retrieval},
  booktitle    = {{EMNLP} (Findings)},
  pages        = {6955--6970},
  year         = {2023}
}

@inproceedings{Pradeep0GLZLM023,
  author       = {Ronak Pradeep and
                  Kai Hui and
                  Jai Gupta and
                  {\'{A}}d{\'{a}}m D. Lelkes and
                  Honglei Zhuang and
                  Jimmy Lin and
                  Donald Metzler and
                  Vinh Q. Tran},
  title        = {How Does Generative Retrieval Scale to Millions of Passages?},
  booktitle    = {{EMNLP}},
  pages        = {1305--1321},
  year         = {2023}
}

@inproceedings{0002DW23,
  author       = {Yujia Zhou and
                  Zhicheng Dou and
                  Ji{-}Rong Wen},
  title        = {Enhancing Generative Retrieval with Reinforcement Learning from Relevance
                  Feedback},
  booktitle    = {{EMNLP}},
  pages        = {12481--12490},
  year         = {2023}
}

@inproceedings{HuangTSG0J0D20,
  author       = {Junjie Huang and
                  Duyu Tang and
                  Linjun Shou and
                  Ming Gong and
                  Ke Xu and
                  Daxin Jiang and
                  Ming Zhou and
                  Nan Duan},
  title        = {CoSQA: 20, 000+ Web Queries for Code Search and Question Answering},
  booktitle    = {{ACL/IJCNLP}},
  pages        = {5690--5700},
  year         = {2021}
}

@inproceedings{LiZYOR24,
  author       = {Zehan Li and
                  Jianfei Zhang and
                  Chuantao Yin and
                  Yuanxin Ouyang and
                  Wenge Rong},
  title        = {ProCQA: {A} Large-scale Community-based Programming Question Answering
                  Dataset for Code Search},
  booktitle    = {{LREC/COLING}},
  pages        = {13057--13067},
  year         = {2024}
}

@inproceedings{LuGRHSBCDJTLZSZ21,
  author       = {Shuai Lu and
                  Daya Guo and
                  Shuo Ren and
                  Junjie Huang and
                  Alexey Svyatkovskiy and
                  Ambrosio Blanco and
                  Colin B. Clement and
                  Dawn Drain and
                  Daxin Jiang and
                  Duyu Tang and
                  Ge Li and
                  Lidong Zhou and
                  Linjun Shou and
                  Long Zhou and
                  Michele Tufano and
                  Ming Gong and
                  Ming Zhou and
                  Nan Duan and
                  Neel Sundaresan and
                  Shao Kun Deng and
                  Shengyu Fu and
                  Shujie Liu},
  title        = {CodeXGLUE: {A} Machine Learning Benchmark Dataset for Code Understanding
                  and Generation},
  booktitle    = {NeurIPS Datasets and Benchmarks},
  year         = {2021}
}

@inproceedings{RobertsonWJHG94,
  author       = {Stephen E. Robertson and
                  Steve Walker and
                  Susan Jones and
                  Micheline Hancock{-}Beaulieu and
                  Mike Gatford},
  title        = {Okapi at {TREC-3}},
  booktitle    = {{TREC}},
  series       = {{NIST} Special Publication},
  pages        = {109--126},
  year         = {1994}
}

@inproceedings{ZhangATDNR0X24,
  author       = {Dejiao Zhang and
                  Wasi Uddin Ahmad and
                  Ming Tan and
                  Hantian Ding and
                  Ramesh Nallapati and
                  Dan Roth and
                  Xiaofei Ma and
                  Bing Xiang},
  title        = {Code Representation Learning at Scale},
  booktitle    = {{ICLR}},
  year         = {2024}
}

@inproceedings{GaoZLYZCZ025,
  author       = {Zuchen Gao and
                  Zizheng Zhan and
                  Xianming Li and
                  Erxin Yu and
                  Haotian Zhang and
                  Bin Chen and
                  Yuqun Zhang and
                  Jing Li},
  title        = {{OASIS:} Order-Augmented Strategy for Improved Code Search},
  booktitle    = {{ACL}},
  pages        = {18451--18467},
  year         = {2025}
}

@inproceedings{LiuMJSXZY25,
  author       = {Ye Liu and
                  Rui Meng and
                  Shafiq Joty and
                  Silvio Savarese and
                  Caiming Xiong and
                  Yingbo Zhou and
                  Semih Yavuz},
  title        = {Code{XE}mbed: A Generalist Embedding Model Family for Multilingual and Multi-task Code Retrieval},
  booktitle    = {{COLM}},
  year         = {2025}
}

@inproceedings{Zeng0JSWZ24,
  author       = {Hansi Zeng and
                  Chen Luo and
                  Bowen Jin and
                  Sheikh Muhammad Sarwar and
                  Tianxin Wei and
                  Hamed Zamani},
  title        = {Scalable and Effective Generative Information Retrieval},
  booktitle    = {{WWW}},
  pages        = {1441--1452},
  year         = {2024}
}

@inproceedings{WangWWWZLWL22,
  author       = {Xin Wang and
                  Yasheng Wang and
                  Yao Wan and
                  Jiawei Wang and
                  Pingyi Zhou and
                  Li Li and
                  Hao Wu and
                  Jin Liu},
  title        = {{CODE-MVP:} Learning to Represent Source Code from Multiple Views
                  with Contrastive Pre-Training},
  booktitle    = {{NAACL-HLT} (Findings)},
  pages        = {1066--1077},
  year         = {2022}
}

@inproceedings{0003R0YZCRR25,
  author       = {Shiguang Wu and
                  Zhaochun Ren and
                  Xin Xin and
                  Jiyuan Yang and
                  Mengqi Zhang and
                  Zhumin Chen and
                  Maarten de Rijke and
                  Pengjie Ren},
  title        = {Constrained Auto-Regressive Decoding Constrains Generative Retrieval},
  booktitle    = {{SIGIR}},
  pages        = {2429--2440},
  year         = {2025}
}

@inproceedings{Tang0GR0C24,
  author       = {Yubao Tang and
                  Ruqing Zhang and
                  Jiafeng Guo and
                  Maarten de Rijke and
                  Wei Chen and
                  Xueqi Cheng},
  title        = {Generative Retrieval Meets Multi-Graded Relevance},
  booktitle    = {NeurIPS},
  year         = {2024}
}

@inproceedings{BuiYJ21,
  author       = {Nghi D. Q. Bui and
                  Yijun Yu and
                  Lingxiao Jiang},
  title        = {Self-Supervised Contrastive Learning for Code Retrieval and Summarization
                  via Semantic-Preserving Transformations},
  booktitle    = {{SIGIR}},
  pages        = {511--521},
  year         = {2021}
}

@inproceedings{DuSWSHZ21,
  author       = {Lun Du and
                  Xiaozhou Shi and
                  Yanlin Wang and
                  Ensheng Shi and
                  Shi Han and
                  Dongmei Zhang},
  title        = {Is a Single Model Enough? MuCoS: {A} Multi-Model Ensemble Learning
                  Approach for Semantic Code Search},
  booktitle    = {{CIKM}},
  pages        = {2994--2998},
  year         = {2021}
}

@inproceedings{ZhangL0DLC24,
  author       = {Peitian Zhang and
                  Zheng Liu and
                  Yujia Zhou and
                  Zhicheng Dou and
                  Fangchao Liu and
                  Zhao Cao},
  title        = {Generative Retrieval via Term Set Generation},
  booktitle    = {{SIGIR}},
  pages        = {458--468},
  year         = {2024}
}

@article{GraziaP23,
  author       = {Luca Di Grazia and
                  Michael Pradel},
  title        = {Code Search: {A} Survey of Techniques for Finding Code},
  journal      = {{ACM} Comput. Surv.},
  volume       = {55},
  number       = {11},
  pages        = {220:1--220:31},
  year         = {2023}
}

@article{XieLDZW24,
  author       = {Yutao Xie and
                  Jiayi Lin and
                  Hande Dong and
                  Lei Zhang and
                  Zhonghai Wu},
  title        = {Survey of Code Search Based on Deep Learning},
  journal      = {{ACM} Trans. Softw. Eng. Methodol.},
  volume       = {33},
  number       = {2},
  pages        = {54:1--54:42},
  year         = {2024}
}

@article{LiJZZZZD25,
  author       = {Xiaoxi Li and
                  Jiajie Jin and
                  Yujia Zhou and
                  Yuyao Zhang and
                  Peitian Zhang and
                  Yutao Zhu and
                  Zhicheng Dou},
  title        = {From Matching to Generation: {A} Survey on Generative Information
                  Retrieval},
  journal      = {{ACM} Trans. Inf. Syst.},
  volume       = {43},
  number       = {3},
  pages        = {83:1--83:62},
  year         = {2025}
}

@inproceedings{SiSCCZ0000G24,
  author       = {Zihua Si and
                  Zhongxiang Sun and
                  Jiale Chen and
                  Guozhang Chen and
                  Xiaoxue Zang and
                  Kai Zheng and
                  Yang Song and
                  Xiao Zhang and
                  Jun Xu and
                  Kun Gai},
  title        = {Generative Retrieval with Semantic Tree-Structured Identifiers and
                  Contrastive Learning},
  booktitle    = {{SIGIR-AP}},
  pages        = {154--163},
  year         = {2024}
}

@inproceedings{DongCL11,
  author       = {Wei Dong and
                  Moses Charikar and
                  Kai Li},
  title        = {Efficient k-nearest neighbor graph construction for generic similarity measures},
  booktitle    = {{WWW}},
  pages        = {577--586},
  year         = {2011}
}

@inproceedings{XuT0YLZL25,
  author       = {Bo Xu and
                  Yicen Tian and
                  Xiaokun Zhang and
                  Erchen Yu and
                  Dailin Li and
                  Linlin Zong and
                  Hongfei Lin},
  title        = {Reinforcement Learning-Driven Generative Retrieval with Semantic-aligned
                  Multi-Layer Identifiers},
  booktitle    = {{CIKM}},
  pages        = {3592--3601},
  year         = {2025}
}

@inproceedings{WeiL17,
  author       = {Huihui Wei and
                  Ming Li},
  title        = {Supervised Deep Features for Software Functional Clone Detection by
                  Exploiting Lexical and Syntactical Information in Source Code},
  booktitle    = {{IJCAI}},
  pages        = {3034--3040},
  year         = {2017}
}

@inproceedings{DongHZZWZJL25,
  author       = {Yanmin Dong and
                  Zhenya Huang and
                  Zheng Zhang and
                  Guanhao Zhao and
                  Likang Wu and
                  Hongke Zhao and
                  Binbin Jin and
                  Qi Liu},
  title        = {Enhancing Code Search Intent with Programming Context Exploration},
  booktitle    = {{WSDM}},
  pages        = {596--605},
  year         = {2025}
}

@inproceedings{LiuZWJXZ00026,
  author       = {Xiaoyu Liu and
                  Fuwei Zhang and
                  Yiqing Wu and
                  Xinyu Jia and
                  Zenghua Xia and
                  Fuzhen Zhuang and
                  Zhao Zhang and
                  Fei Jiang and
                  Wei Lin},
  title        = {CAT-ID\({}^{\mbox{2}}\): Category-Tree Integrated Document Identifier
                  Learning for Generative Retrieval In E-commerce},
  booktitle    = {{WSDM}},
  pages        = {426--435},
  year         = {2026}
}

@inproceedings{LeeKKCH22,
  author       = {Doyup Lee and
                  Chiheon Kim and
                  Saehoon Kim and
                  Minsu Cho and
                  Wook{-}Shin Han},
  title        = {Autoregressive Image Generation using Residual Quantization},
  booktitle    = {{CVPR}},
  pages        = {11513--11522},
  year         = {2022}
}

@article{ChenWCDW23,
  author       = {Yuhong Chen and
                  Zhihao Wu and
                  Zhaoliang Chen and
                  Mianxiong Dong and
                  Shiping Wang},
  title        = {Joint learning of feature and topology for multi-view graph convolutional
                  network},
  journal      = {Neural Networks},
  volume       = {168},
  pages        = {161--170},
  year         = {2023}
}

@article{WuLLCBW23,
  author       = {Zhihao Wu and
                  Xincan Lin and
                  Zhenghong Lin and
                  Zhaoliang Chen and
                  Yang Bai and
                  Shiping Wang},
  title        = {Interpretable Graph Convolutional Network for Multi-View Semi-Supervised
                  Learning},
  journal      = {IEEE Transactions on Multimedia},
  volume       = {25},
  pages        = {8593--8606},
  year         = {2023}
}

@article{FengLWWXWL25,
  author       = {Jiadong Feng and
                  Wei Li and
                  Suhuang Wu and
                  Zhao Wei and
                  Yong Xu and
                  Juhong Wang and
                  Hui Li},
  title        = {Deep Code Search with Naming-Agnostic Contrastive Multi-View Learning},
  journal      = {{ACM} Trans. Knowl. Discov. Data},
  volume       = {19},
  number       = {6},
  pages        = {1--26},
  year         = {2025},

}

@article{RQ-MoE,
  author       = {Zhengjia Zhong and
                  Shuyan Ke and
                  Zaizhou Lin and
                  Jiaqi Song and
                  Hongyi Lan and
                  Hui Li},
  title        = {RQ-MoE: Residual Quantization via Mixture of Experts for Efficient
                  Input-Dependent Vector Compression},
  journal      = {arXiv Preprint},
  url          = {https://arxiv.org/abs/2605.14359},
  year         = {2026},
}

\end{document}